# Multiresolution analysis (discrete wavelet transform) through Daubechies family for emotion recognition in speech.

**D Campo[1,2], O L Quintero[2], M Bastidas[2]**

[1] Dipartimento di Ingegneria Navale, Elettrica, Elettronica e delle Telecomunicazioni (DITEN). Information and Signal Processing for Cognitive Telecommunications ISIP40 Italy Genova.

[2] Mathematical Modeling Research Group at Mathematical Sciences Department in School of Sciences at Universidad EAFIT, Medellín Colombia.

Emails: dcampoc@eafit.edu.co , oquinte1@eafit.edu.co, mbastida@eafit.edu.co

**Abstract**. We propose a study of the mathematical properties of voice as an audio signal. This work includes signals in which the channel conditions are not ideal for emotion recognition. Multiresolution analysis- discrete wavelet transform – was performed through the use of Daubechies Wavelet Family (Db1-Haar, Db 6, Db8, Db10) allowing the decomposition of the initial audio signal into sets of coefficients on which a set of features was extracted and analyzed statistically in order to differentiate emotional states. ANNs proved to be a system that allows an appropriate classification of such states. This study shows that the extracted features using wavelet decomposition are enough to analyze and extract emotional content in audio signals presenting a high accuracy rate in classification of emotional states without the need to use other kinds of classical frequency-time features. Accordingly, this paper seeks to characterize mathematically the six basic emotions in humans: boredom, disgust, happiness, anxiety, anger and sadness, also included the neutrality, for a total of seven states to identify.

## 1. Introduction

Emotions have been studied by different knowledge areas; among them are medicine, psychology, neuroscience and the social sciences. An emotional state can be evidenced in the human physiological alteration. In this sense, there are scientific studies that use different measures in order to identify emotional changes in a person; it is the case of brain potentials, facial micro expressions, and galvanic skin conductance together with breathing and blood pressure. In addition to the physiological signals mentioned above, it has been found that emotions can be analyzed by voice signals. In order to detect emotions in audio signals, some scientific researches use traditional features extracted from the primary signal, among them are the formant frequencies, fundamental frequency, signal energy in time, signal spectrum by Fast Fourier Transform (FFT) and the Mel Frequency Cepstral Coefficients (MFCCs). The researches that consider these types of features usually have some limitations when the gender of the speaker is considered [1-18]. Those results show that this kind of features are speaker dependent, this causes loss of generality when detecting emotions. In addition, predictive models based on traditional voice characteristics generally not exceed 70% accuracy on complete databases which consider several speakers and the six basic human emotions: boredom, disgust, happiness, anxiety, anger and sadness. There is a variety of researches that use traditional features as inputs of inference systems based on fuzzy logic, Artificial Neural Networks (ANN), Support Vector Machine (SMV), K-nearest neighbors (KNN), Hidden Markov Models (HMM) and Gaussian Mixture Models (GMM) [19-21]. In addition to the traditional classification features, there are some other scientific papers [22-24] that use a signal decomposition based on the wavelet transform in order to extract features that provide higher results in terms of speaker independence and overall accuracy percentage in the recognition of complete emotional databases. Those results demonstrate that the use of the wavelet decomposition technique has advantages over the spectral analysis that can be done using the Fourier transform. One of the most important advantages of wavelet decomposition is that it allows extracting information from the signal to be analyzed. In that way, wavelets act as filters that allow the extraction of high and low frequencies from an initial signal. Depending on the wavelet that is used; the filter acting over the audio signal will have certain properties. In this vein, the present article seeks to employ selected wavelets to distinguish certain features of an emotion. There have been identified some scientific studies in which wavelet decomposition is used to recognize emotions over speech signals. In the first [25] is considered a database of emotions from 5 native languages of Assam, they use an ANN classifier. Another research considers a database in Mandarin and they use a HMM classifier [27]. Additionally, there is a work [26] in which the wavelet transform is used in conjunction with some statistical techniques on the Berlin database.

The studies that involve wavelet analysis in speech emotion recognition (SER) have an accuracy of over 80%; and they use wavelet decomposition together with other techniques that allow differentiating emotional states in speech. In addition, these studies usually tend windowing the audio signals [24], [27], [28] in order to find features that allow detecting emotions in short speech expressions.

This paper presents an analysis of emotions in audio signals using wavelet transform without another kind of features. It is hypothesized that wavelet decomposition is sufficient for classifying emotions in audio signals. Also, the audio signals are analyzed without using windows over the initial signal, thus, it is possible to find global voice characteristics. Finally, this article also seeks to accomplish the construction of an inference system based on artificial intelligence that uses wavelet decomposition features as inputs. This work was performed using the Berlin Emotional Speech Database (Emo-DB), which has proven to be reliable in the emotional content which is reflected in its audio signals [29].

The paper is organized as follows. In Section 2 there is a review of the essentials issues of wavelet transform and it is also shown the principal properties of the wavelet family used in this work. In section 3 is shown the proposed methodology for emotion identification in audio signals. Section 4 shows the results obtained using the proposed method and finally some conclusions are exposed in section 5.

## 2. Multiresolution analysis

2.1. *Theoretical aspects of discrete wavelet transform*
Let $\psi(x)$ be a mother wavelet, the operations of translation and dilation are defined by the following expression:

$$\psi_{jk}(x) = 2^{j/2}\psi(2^j x - k) \quad (1)$$

Where $\psi_{jk}(x)$, for all j and k forms an orthonormal basis for $L^2(R)$ function. This space, for fixed j is defined by the following expression:

$$V_j = \left\{\sum_{k=-\infty}^{\infty} a_k \psi_{jk}(x) / \sum_k |a_k|^2 < \infty\right\} = n\left\{\psi_{jk}(x), k \in Z\right\} \quad (2)$$

A function or signal $f(x) \in L^2(R)$ can be expressed by the form

$$f(x) = \sum_{j=-\infty}^{\infty}\sum_{k=-\infty}^{\infty} \alpha_{jk}\psi_{jk}(x) \quad (3)$$

The $α_{jk}$ are the wavelet transform coefficients of f(x). Every orthonormal mother has an auxiliary wavelet function $ϕ(x) ∈ L^2(R)$ called father wavelet or scaling wavelet function and satisfies the following properties:

$$\int_{-\infty}^{\infty} ϕ(x)dx = 1$$
$$\|ϕ(x)\|^2 = \int_{-\infty}^{\infty}|ϕ(x)|^2 dx = 1 \quad (4)$$
$$⟨ϕ(x)ϕ(x-n)⟩ = \int_{-\infty}^{\infty} ϕ(x)ϕ(x-n)dx = δ(n)$$

$V_0$ is defined as the subspace of $L^2(R)$ generated by the set of moved scale functions $ϕ_k(x)$, such that:

$$V_0 = gen\{ϕ_k(x) = ϕ(x-k), k ∈ Z\} \quad (5)$$

Then any function $V_0$ can be expressed as a linear combination of base functions, that is to say, a function $f(x) ∈ L^2(R)$ can be projected onto the subspace $V_0$ so that the development of this projection in terms of the moved scale functions is an approximation of f(x) such that $P_0 f = f_0(x) = \sum_{-\infty}^{\infty} a_k ϕ_k(x)$, where $a_k = ⟨f(x)ϕ_k(x)⟩$. In general, closed subspaces $V_j$ are obtained:

$$V_j = gen\left\{ϕ_{jk}(x) = 2^{\frac{j}{2}}ϕ(2^j x - k), k, j ∈ Z\right\} \quad (6)$$

Thus, if $f(x) ∈ L^2(R)$ can be projected onto the subspaces $V_j$ so that the development of such a projection in terms of moved scale functions constitutes an approximation of f(x), such that $P^j f = f_j(x) = \sum_{-\infty}^{\infty} a_k ϕ_{jk}(x)$, where $a_k = ⟨f(x)ϕ_{jk}(x)⟩$. $V_j$ subspaces are associated with scale function $ϕ(x)$, where the effect of the indices k and j on $ϕ(x)$ is as follows: k moves the function along x axis and j compresses in $2^j$ the function along the x axis. Once projected a function or signal f(x) in the $V_j$ subspaces, an approximation of f(x) is obtained and relevant information associated to frequency details of f(x) is missed [35]. In this vein, $W_j$ subspaces are defined as the orthogonal complement of $V_j$ in $V_{j+1}$. In this way, f(x) can be rewritten in terms of an approximation subspace $V_j$ in the j resolution and the remaining orthogonally terms containing the finer details associated with the subspace $W_j$ associated with the mother wavelet ψ(x). Apply the Wavelet Packet Decomposition (WPD) on a signal is equivalent to divide it into a group of approximation coefficients ($V_j$) and detail coefficients ($W_j$), where j defines the number of times this decomposition process is performed.

*2.2. Daubechies Family of Wavelets*

Wavelet packets were introduced by Coifman, Meyer and Wickerhauser [30] by generalizing the link between multiresolution approximations and wavelets [31]. Accordingly, a wavelet is a signal (or waveform) with limited duration whose mean value is zero. The wavelets tend to be irregular and asymmetric. The analysis by the Discrete Wavelet Packet Transform (DWPT) consists in decomposing a signal into scaled functions transferred in time from an original wavelet, called mother wavelet [32]. Signals with sharp changes are best analyzed by irregular wavelets [33]. Such wavelets are empowered to analyze local areas of large signals. For the calculation of the wavelet transform a mother wavelet must be chosen and it will serve as a prototype for all windows that are used in the process. There is an important collection of wavelet functions that have proved to be extremely useful, among which are: Haar, Daubechies, Biorthogonal, Coiflets, Symetrics, Morlet, Mexican Hat and Meyer. In order to carry out a WPD over a signal the wavelet collection should form an orthonormal base which relies on the concept of multi-resolution analysis [34]. In this sense, this paper uses 4 orthonormal wavelets in order to perform a WPD on audio signals from the Emo-DB.

This work involves an analysis over audio signals in which is possible to take advantage of the properties of those signals in order to perform a methodology which can deal with the characterization of the emotional content into them. As is well known, human voice has a wave nature, more specifically, the speech is basically generated when sound pressure waves are channeled or restricted in various ways by manual control of the anatomical components of the human vocal system [32]. Base on that, in this paper is considered a wavelet analysis over the audio signals exploiting the properties they have to express periodic signals into decomposed new signals. Wavelet analysis has proven to be a better approach than Fourier transforms at the time to characterize particular emotions in speech signals [25], [26] and [27].

## 3. Statistical features extraction and neural networks classifiers

### 3.1. Statistical features extraction

As is well known, statistics and probability are used in Digital Signal Processing to characterize signals and processes that generate them [30]. In the treatment of signals is very common to take into consideration high order statistics to extract information about the signal. The most conventional techniques involve second order statistics and the signal spectrum [31].

This work uses some statistical measurements over the decomposed signals in order to extract relevant features that allow recognizing the speaker emotion through the use of the Daubechies wavelet family. The principal statistical measurements extracted are: Maximum values, mean, median, interquartile range, standard deviation, zero frequency crossings, kurtosis, sknewness. The statistical features were extracted for different levels of wavelet decomposition. The statistical signal classification is an important component for the categorization of input data into identifiable digital pattern classes using extraction significant features [33], those statistical methods have been used in many areas of signal processing, including speech recognition [34]. In the present proposal is used a statistical hypothesis test for mean difference in order to separate two classes, each emotion represents a class and it is considered all the possible pairs among the 7 emotional states to perform the test.

### 3.2. Artificial Neural Network Classifier

The use of ANNs as classifiers in speech recognition problems has been considered by researchers due to the accuracy that this method can present when it is implemented correctly and the capacity that it has to generalize the information responding appropriately to inputs that differ from those with which the training step of the network was done [35]. The use of ANN has been proved to have better performance than other methods in the classification of emotions over speech signals considering both frameworks: speaker dependent and independent focuses [36]. It represents a motivation to choose an ANN as an emotion classifier. It was identified a work [37] that proposes a set of Bayesian classifiers considering every pair of emotions over 5 principal states (anger, happiness, neutral, sadness and surprise) in Japanese speech. A pairwise classification of emotions is performed in that article.

In the present work a pairwise selection of emotions is performed; this focus allows solving a multi-class problem using a binary approach in which the final features are joined in a final neural network structure whose outputs are the membership value to each emotion of interest.

## 4. Methodology

First, all the audios were decomposed in ten levels using the wavelets Db1, Db6, Db8 and Db10. A big set of features was extracted in each level and a statistical hypothesis testing was performed for each pair of emotions in order to select the most relevant features that allow to differentiate the emotions. Specifically the test performed was the Student's t-test, which is used to determine if two set of data are different from each other. In some sense, this process was executed in order to do a projection of the big set of features into a littler one.

*Figure 1. Methodology for the decomposition and feature selection*

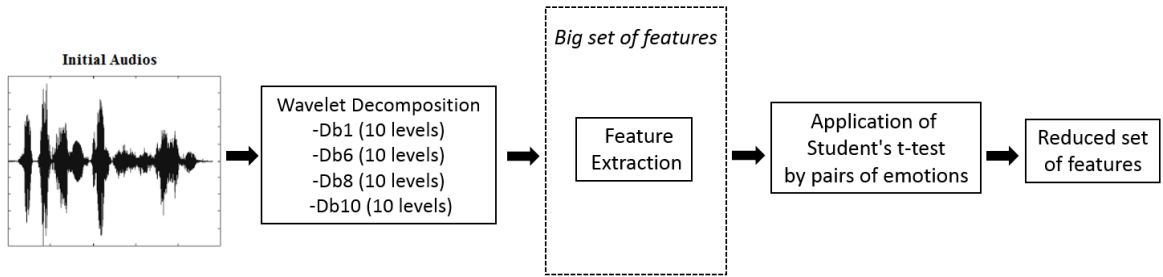

A simple ANN was trained based on the new set of features selected by the statistical hypothesis method. The final ANN structure used in this work is as follows: the output layer has 7 neurons due to the total numbers of emotional states. The number of inputs is 42 due to the process that was performed using the statistical hypothesis testing, in which it was assumed that two features for each pair of emotions are enough to classify the seven emotions of the database. Finally, it was considered only one hidden layer with 50 neurons whose transfer function is the hyperbolic tangent sigmoid.

*Figure 2. Final architecture of the used ANN*

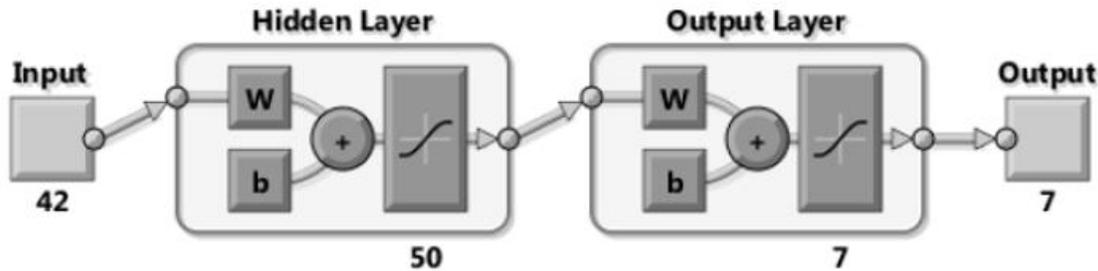

The assumption that is taken into account in this work is based on the fact that is possible to identify a specific emotion inside a group if we are able to classify correctly each possible pair of emotion into the group.

## 5. Results

The confusion matrix for the first proposed ANN is shown in the table 1. From the results, it can be seen that the most difficult emotional state to detect among the seven proposed is happiness with 85.71%, while the easiest one to detect is sadness with 93.33%.

Thus, the Table 2 shows the error rates per couple of emotions was obtained. From the table, it can be seen that the higher rate of error occurs when trying to differentiate audio signals associated to boredom and neutrality. This was expected due to the selection of the features that was done in the beginning, where it was observed that features associated with the detection of the pair Boredom-Neutral had a lower percentage of accuracy. It can be seen in Table 2 there are five pairs of emotions are classified with 100% accuracy, i.e. with 0% error, these pairs of emotional states are Boredom-Happiness, Disgust-Sadness, Happiness-Sadness, Neutral-Anger and Anger-Sadness.

If the results of the confusion matrices there is that neutrality is detected with the same precision in both cases. Boredom, happiness and anxiety are better classified by the first proposal Consequently the total accuracy for this proposal is 90.96%.

Table 1. Confusion matrix for the first ANN proposal

| Emotional states | Boredom | Disgust | Happiness | Anxiety | Neutral | Anger | Sadness |
|---|---|---|---|---|---|---|---|
| **Boredom** | 89.74% | 00.00% | 00.00% | 01.28% | 06.41% | 01.28% | 01.28% |
| **Disgust** | 02.22% | 91.11% | 04.44% | 00.00% | 02.22% | 00.00% | 00.00% |
| **Happiness** | 00.00% | 02.86% | 85.71% | 02.86% | 01.43% | 07.14% | 00.00% |
| **Anxiety** | 01.45% | 01.45% | 01.45% | 92.75% | 02.9% | 00.00% | 00.00% |
| **Neutral** | 06.25% | 00.00% | 00.00% | 02.5% | 91.25% | 00.00% | 00.00% |
| **Anger** | 00.00% | 00.79% | 04.76% | 01.59% | 00.00% | 92.86% | 00.00% |
| **Sadness** | 01.67% | 00.00% | 00.00% | 01.67% | 03.33% | 00.00% | 93.33% |

Table 2. Error percentage by pair of emotional states using the first ANN proposal.

| Pair of emotional states | Error percentage |
|---|---|
| Boredom - Disgust | 1.11% |
| Boredom - Happiness | 0.00% |
| Boredom - Anxiety | 1.36% |
| Boredom - Neutral | 6.33% |
| Boredom - Anger | 0.64% |
| Boredom - Sadness | 1.47% |
| Disgust - Happiness | 3.65% |
| Disgust - Anxiety | 0.72% |
| Disgust - Neutral | 1.11% |
| Disgust - Anger | 0.39% |
| Disgust - Sadness | 0.00% |
| Happiness - Anxiety | 2.15% |
| Happiness - Neutral | 0.71% |
| Happiness - Anger | 5.95% |
| Happiness - Sadness | 0.00% |
| Anxiety - Neutral | 2.75% |
| Anxiety - Anger | 0.79% |
| Anxiety - Sadness | 0.83% |
| Neutral - Anger | 0.00% |
| Neutral - Sadness | 1.66% |
| Anger - Sadness | 0.00% |

## 6. Conclusions

Feature's extraction from voice as audio signals in no ideal conditions is a challenge for emotion recognition purposes. Wavelet approximation through multi resolution analysis, results to be an accurate methodology for decomposition of the signal in several levels so called approximation and details. The use of statistical measurements over this signal decomposition for features selection, allows the recognition of several emotional states through the use of an Artificial Neural Network as classifier.

Future work will present this approach over audio signals for real applications.